# Information Hiding Using Improper Frame Padding

Bartosz Jankowski, Wojciech Mazurczyk, Krzysztof Szczypiorski

*Abstract* — Hiding information in network traffic may lead to leakage of confidential information. In this paper we introduce a new steganographic system: the *PadSteg* (Padding Steganography). To authors' best knowledge it is the first information hiding solution which represents *interprotocol steganography* i.e. usage of relation between two or more protocols from the TCP/IP stack to enable secret communication. *PadSteg* utilizes ARP and TCP protocols together with an *Etherleak* vulnerability (improper Ethernet frame padding) to facilitate secret communication for hidden groups in LANs (Local Area Networks). Basing on real network traces we confirm that *PadSteg* is feasible in today's networks and we estimate what steganographic bandwidth is achievable while limiting the chance of disclosure. We also point at possible countermeasures against *PadSteg*.

*Keywords: steganography, ARP, frame padding, Etherleak*

## I. INTRODUCTION

Network steganography is currently seen as a rising threat to network security. Contrary to typical steganographic methods which utilize digital media (pictures, audio and video files) as a cover for hidden data (steganogram), network steganography utilizes communication protocols' control elements and their basic intrinsic functionality. As a result, such methods may be harder to detect and eliminate.

In order to minimize the potential threat to public security, identification of such methods is important as is the development of effective detection (steganalysis) methods. This requires both an in-depth understanding of the functionality of network protocols and the ways in which it can be used for steganography. Many methods had been proposed and analyzed so far – for the detailed review see Zander et al. [2] or Petitcolas et al. [3].

Typical network steganography method uses modification of a single network protocol. The protocol modification may be applied to the PDU (Protocol Data Unit) [1], [4], [5], time relations between exchanged PDUs [6], or both [14] (hybrid methods). As far as the authors are aware, *PadSteg* (Padding Steganography), presented in this paper, is the first steganographic system that utilizes what we have defined as *interprotocol* steganography i.e. usage of relation between two or more different network protocols to enable secret communication – *PadSteg* utilizes Ethernet (IEEE 802.3), ARP and TCP protocols.

This work was supported in part by the Polish Ministry of Science and Higher Education under Grant: N517 071637.

The authors are with Institute of Telecommunications, Warsaw University of Technology, Nowowiejska 15/19, 00-665 Warsaw, Poland, email: B.Jankowski@stud.elka.pw.edu.pl, wmazurczyk@tele.pw.edu.pl, ksz@tele.pw.edu.pl

ARP (Address Resolution Protocol) [10] is a simple protocol which operates between the data link and network layers of the OSI (Open Systems Interconnection) model. In IP networks it is used mainly to determine the hardware MAC (Media Access Control) address when only a network protocol address (IP address) is known. ARP is vital for proper functioning of any switched LAN (Local Area Network) although it can raise security concerns e.g. it may be used to launch an ARP Poisoning attack.

In Ethernet, frame length is limited to a minimum of 64 octets, due to the CSMA/CD (Carrier Sense Multiple Access/ Collision Detection) mechanism, and a maximum of 1500 octets. Therefore, any frames whose length is less than 64 octets have to be padded with additional data. The minimal size of an Ethernet data field is 46 octets and can be filled with data originating from any upper layer protocol, without encapsulation via the LLC (Link Layer Control), because LLC (with its 8 octets header) is very rarely utilized in 802.3 networks.

However, due to ambiguous standardization (RFC 894 and RFC 1042), implementations of padding mechanism in current NICs (Network Interface Cards) drivers vary. Moreover, some drivers handle frame padding incorrectly and fail to fill it with zeros. As a result of memory leakage, an Ethernet frame padding may contain portions of kernel memory. This vulnerability is discussed in *Atstake* report and is called *Etherleak* [9]. Data inserted in padding by *Etherleak* is considered unlikely to contain any valuable information; therefore it does not pose serious threat to network security as such. However, it creates a perfect candidate for a carrier of the steganograms, thus it may be used to compromise network defenses. Utilization of padding in Ethernet frames for steganographic purposes was originally proposed by Wolf [13]. If every frame has padding set to zeros (as stated in standard), its usage will be easy to detect. With the aid of *Etherleak,* this information hiding scheme may become feasible as it will be hard to distinguish frames affected by *Etherleak* from those with steganogram.

In this paper we propose a new steganographic system *PadSteg*, which can be used in LANs and utilizes ARP and TCP protocols together with an *Etherleak* vulnerability. We conduct a feasibility study for this information hiding system, taking into account the nature of todays' networks. We also suggest possible countermeasures against *PadSteg*.

The rest of the paper is structured as follows. Section 2 describes the *Etherleak* vulnerability and related work with regard to the application of padding for steganographic purposes. Section 3 includes a description of *PadSteg* components. Section 4 presents experimental results for real-life LAN traffic which permit for an evaluation of feasibility of the proposed solution. Section 5 discusses possible methods of detection and/or elimination of the proposed information hiding system. Finally, Section 6 concludes our work.

## II. RELATED WORK

### A. The Etherleak vulnerability

The aforementioned ambiguities within the standardization cause differences in implementation of the padding in Ethernet frames. Some systems have an implemented padding operation inside the NIC hardware (so called *auto padding*), others have it in the software device drivers or even in a separate layer 2 stack.

In the *Etherleak* report Arkin and Anderson [9] presented in details an Ethernet frame padding information leakage problem. They also listed almost 50 device drivers from Linux 2.4.18 kernel that are vulnerable.

Due to the inconsistency of padding content of short Ethernet frames (its bits should be set to zero but in many cases they are not), information hiding possibilities arise. That is why it is possible to use the padding bits as a carrier of steganograms.

Since Arkin and Anderson's report dates back to 2003, we performed an experiment in order to verify whether *Etherleak* is an issue in today's networks. The achieved results confirmed that many NICs are still vulnerable (see experimental results in Section 4).

### B. Data hiding using padding

Padding can be found at any layer of the OSI RM, but typically it is exploited for covert communications only in the data link, network and transport layers.

Wolf in [13], proposed a steganographic method which utilizes padding of 802.3 frames. Its achievable steganographic bandwidth is up to 45 bytes/frame.

Fisk et al. [7] presented padding of the IP and TCP headers in the context of active wardens. Each of these fields offers up to 31 bits/packet for steganographic communication.

Padding of IPv6 packets for information hiding was described by Lucena et al. in [8] and offers a couple of channels with a steganographic bandwidth up to 256 bytes/packet.

## III. COMPONENTS OF THE PROPOSED STEGANOGRAPHIC SYSTEM

*PadSteg* enables secret communication in a hidden group in a LAN environment. In such group, each host willing to exchange steganograms should be able to locate and identify other hidden hosts. To provide this functionality certain mechanisms must be specified. In our proposal, ARP protocol, together with improper Ethernet frame padding are used to provide localization and identification of the members of a hidden group. To exchange steganograms improper Ethernet frame padding is utilized in frames that in upper layer use TCP protocol. That is why, in this section we first describe ARP protocol, and then we focus on proposed steganographic system operations.

### A. Overview of ARP Protocol

ARP returns the layer 2 (data link) address for a given layer 3 address (network layer). This functionality is realized with two ARP messages: Request and Reply. The ARP header is presented in Fig. 1.

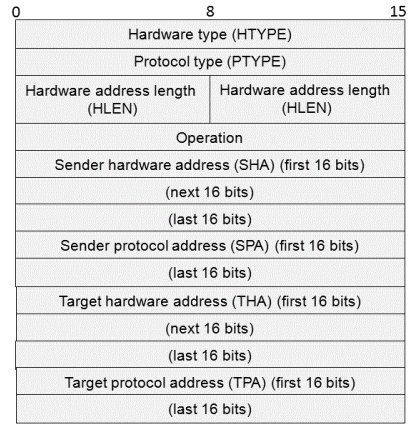

Figure 1. ARP header format

ARP header fields have the following functions:
- HTYPE (Hardware Type) – type of data link protocol used by sender (1 is inserted if it is Ethernet).
- PTYPE (Protocol Type) – type of network protocol in network layer (0800h is inserted if IP is used).
- HLEN (Hardware Length) – length of hardware address fields: SHA, THA (in bytes).
- PLEN (Protocol Length) – length of protocol address fields: SPA, THA (in bytes).
- OPER (Operation) – defines, whether the frame is an ARP REQUEST (1) or REPLY (2) message.
- SHA (Sender Hardware Address) – sender data link layer address (MAC address for Ethernet).
- SPA (Sender Protocol Address) – sender network layer address.
- THA (Target Hardware Address) – data link layer address of the target. This field contains zeros whenever a REQUEST ARP message is sent.
- TPA (Target Protocol Address) – network layer address of the target. This field contains zeros if REQUEST ARP message is sent.

An example of ARP communication with Request/Reply exchange, captured with the *Wireshark* sniffer (www.wireshark.org), is presented in Fig. 2. First, ARP Request is issued (1), which is used by the host with IP address 10.7.6.29 to ask other stations (by means of broadcast): 'Who has IP 10.7.56.47?'. In order to send a frame intended for everyone in a broadcast domain, Ethernet header destination address must be set to FF:FF:FF:FF:FF:FF (2). Next, host with IP address

10.7.56.47 replies directly to 10.7.6.29 using unicast ARP Reply (3) with its MAC address.

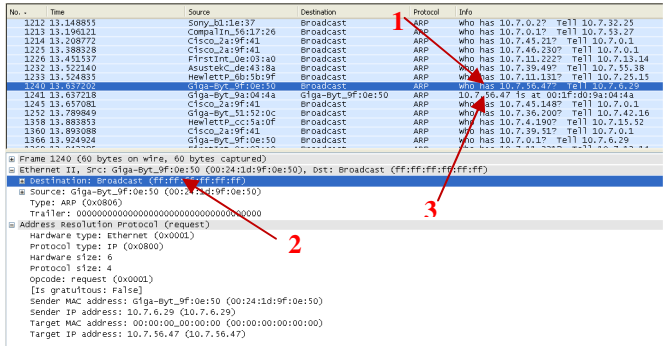

**Figure 1. ARP exchange captured with *Wireshark***

Basing on the proposed description of ARP protocol, it can be concluded that ARP header is rather of fixed content and presents little possibilities for information hiding. One opportunity is to modulate address fields like it was proposed in [11] or [8]. However, this solution provides limited steganographic bandwidth if certain level of undetectability is to be achieved. Moreover, it may result in improper IP and MAC address advertisements which may make this method more prone to detection.

Thus, in the proposed steganographic system *PadSteg*, we utilize ARP Request messages, broadcasted throughout LAN, to make other members of the hidden group become aware of the presence of a new member.

*B. Steganographic system operation*

*PadSteg* is designed for LANs only because it utilizes improper Ethernet frame padding in Ethernet. It also uses ARP and TCP protocols to control hidden groups and steganograms exchange. *PadSteg* operation can be split into two phases:
- Phase I: Advertisement of the hidden nodes.
- Phase II: Hidden data exchange.

*Phase I*

This phase is based on the exchange of ARP Request messages with improper Ethernet frame padding (Fig. 3).

Hidden node that wants to advertise itself to others in the group, broadcasts an ARP Request message (1) and inserts *advertising sequence* into the padding bits. It consists of: a random number (different from 0) and hash which is calculated based on this random number and source MAC address. An example of the padding bits format (which for ARP is 144 bits long), for the MD5 hash function, is presented in Fig. 4.

All the hidden nodes are obligated to analyze the padding of all received ARP Requests. If an ARP Request is received with padding that is not all zeros, it is analyzed by extracting the random number and the corresponding hash is calculated (2). If the received and calculated hashes are the same then it means that a new hidden node is available for steganographic exchange. Each hidden node stores a list of nodes from which it has received advertisements. Every node should also reissue ARP Request at certain time intervals to inform other hidden nodes about its existence. To limit the chance of detection, sending of ARP Requests may not happen too often (3, 4). In ARP, if an entry in host ARP cache is not refreshed within 1 to 20 minutes (implementation dependent) it expires and is removed. Thus, hidden nodes should mimic such behavior to imitate the sending of ARP Requests caused by ARP cache expiration.

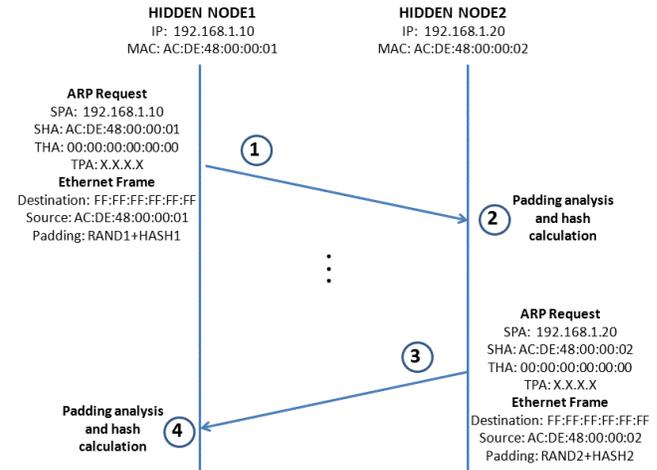

**Figure 3. Hidden group activation phase**

Adoption of ARP messages for identification of new hidden nodes has two advantages:
- The broadcast messages will be received by all hosts in LAN.
- The ARP traffic totals to about 0.1% of all traffic (see next Section for details), so this choice is also beneficial from the performance perspective. Each hidden node does not have to analyze all of the received traffic but only ARP Requests.

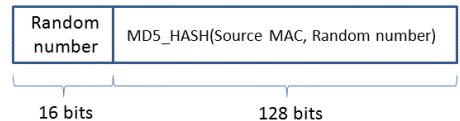

**Figure 4. Padding format of ARP Request messages for the activation phase**

*Phase II*

After the identification of a new hidden node other hidden nodes analyze Ethernet frame padding in every TCP segment sent from that MAC address. The received TCP segments' padding contains steganogram bits.

The bidirectional transmission is realized as presented in Fig. 5. Two hidden nodes make a overt TCP connection e.g. they transfer a file (1). During the connection TCP ACK segments are issued with improper Ethernet frame padding (2 and 4). Received TCP segments are analyzed for improper Ethernet padding presence and secret data is extracted (3 and

5). For third party observer such communication looks like usual data transfer.

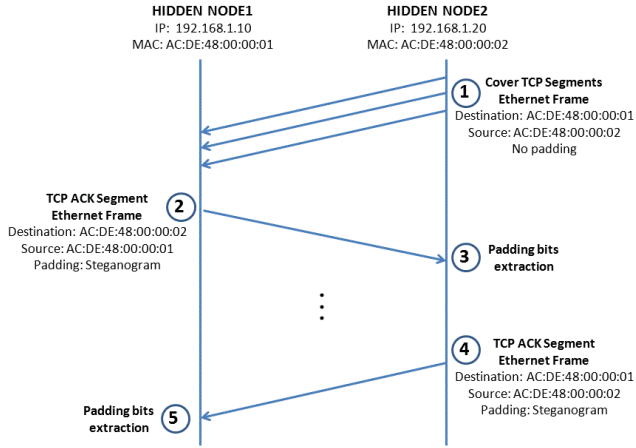

**Figure 5. Hidden group steganograms exchange phase**

## IV. EXPERIMENTAL RESULTS AND STEGANOGRAPHIC BANDWIDTH ESTIMATION

### A. ARP and improper Ethernet frame padding in real-life networks

To evaluate the proposed steganographic system: its steganographic bandwidth and undetectability, real network traffic from LAN was captured.

The experiment was conducted at the Institute of Telecommunications at Warsaw University of Technology between 15 and 19 of March 2010 (from Monday to Friday). It resulted in about 37 million packets captured, which corresponds, daily, to 7.43 million frames on average (with a standard deviation 1.2 million frames) – for details see Table 1. The traffic was captured with the aid of *Dumpcap* which is part of the *Wireshark* sniffer ver. 1.3.3 (www.wireshark.org). The sources of traffic were ordinary computer devices placed in several university laboratories and employees' ones but also peripherals, servers and network equipment. To analyze the captured traffic and calculate statistics *TShark* (which is also part of *Wireshark*) was utilized. Statistics were calculated per day, and average results are presented.

TABLE I. THE NUMBER OF CAPTURED FRAMES PER DAY

| *Date* | *Monday* | *Tuesday* | *Wednesday* | *Thursday* | *Friday* |
|---|---|---|---|---|---|
| No. of frames | 7,205,904 | 7,027,170 | 5,761,723 | 8,241,832 | 8,945,403 |

The captured traffic classification by upper layer protocol is presented in Fig. 6. Three quarters of the traffic was HTTP. Together with SSH, UDP and SSL protocols it sums up to about 93% of the traffic.

Almost 22% (with a standard deviation of 7.7%) of all daily traffic had padding bits added (~8 million frames). It is obvious that not all of the frames were affected since padding is added only to small-sized packets.

The upper layer protocols that most frequently led to generation of padded frames are presented in Fig. 7.

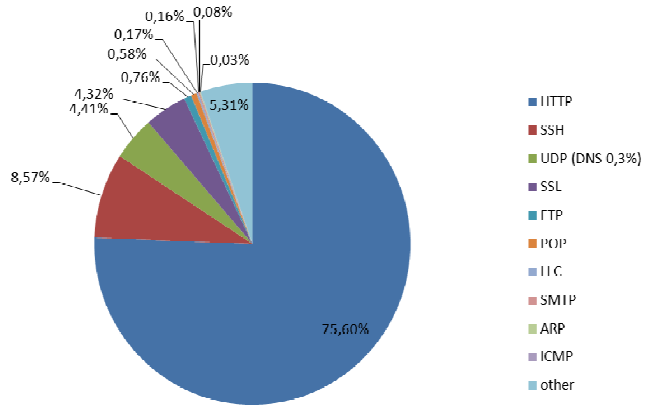

**Figure 6. Captured traffic characteristics**

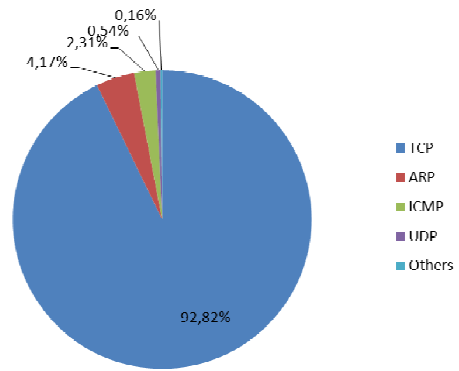

**Figure 7. Upper layer protocols that result in padded Ethernet frames**

However, it is important to note, that almost 22% of the padded frames experienced improper padding (~1.8 million frames). These frames were generated by about 15% of hosts in the inspected network (their NICs were produced among others by some US leading vendors). We considered Ethernet frame padding improper if the padding bits were not set to zeros. Table 2 shows for which network protocols frames were mostly improperly padded.

TABLE II. UPPER LAYER PROTCOLS AFFECTED WITH ETHERNET FRAME IMPROPER PADDING IN EXPERIMENTAL DATA

| *Affected protocol* | *TCP* | *ARP* | *ICMP* | *Others* |
|---|---|---|---|---|
| [%] | 93.19 | 4.17 | 2.31 | 0.32 |

TCP segments with an ACK flag set (which have no payload) result in frames that have to be padded, thus, it is no surprise that ~93% of improperly padded traffic is TCP. Nearly all of this traffic consists of ACK segments. Other frames that had improper padding were caused by ARP and ICMP messages (~6.5%).

*PadSteg* is based on the ARP protocol, thus our aim was also to find out ARP statistics i.e. what are the most frequently used ARP messages, what is their distribution and how many of them have improper padding. The results are presented in Fig. 8.

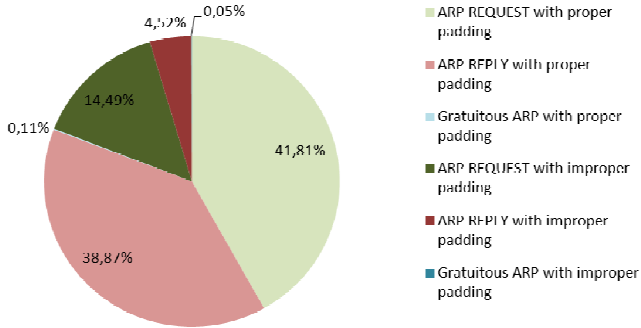

**Figure 8. Captured ARP characteristics**

Not surprisingly, the most frequently sent ARP messages were ARP Request (~56.3%) and Reply (~43.4%), while Gratuitous ARP messages are in minority (~0.2%). Out of all ARP messages almost 20% had improper padding.

*B. Steganographic bandwidth estimation*

Let us try to estimate *PadSteg* steganographic bandwidth for a single hidden node transmitting in a hidden group.

Because, currently, there are no tools for steganography detection, in real-life networks, every member of a hidden group can exchange almost unlimited number of steganograms and remain undiscovered. However, if the network traffic is consequently monitored, a naive use of *PadSteg* – that is: excessive generation of Ethernet frames with improper padding may be easily detected.

This leads to conclusion that it is important to evaluate what is the realistic steganographic bandwidth under the assumption that the secret data exchange will not differ from other hosts' traffic burdened with the *Etherleak* vulnerability. To achieve this goal steganographic user's network activity must mimic behavior of other users in terms of sending Ethernet frames with improper padding.

We calculated the steganographic bandwidth of the proposed system based on the average, daily number of TCP segments with improper Ethernet padding per susceptible host (see Table 3).

Because each TCP segment's padding is 6 bytes long, the average steganographic bandwidth is about 27 bit/s (with a standard deviation of about 12 bit/s). Therefore, if the hidden node generates Ethernet frames with improper padding that fall within the average range, for the inspected LAN network, steganographic communication may remain undetected.

TABLE III. THE NUMBER OF FRAMES WITH IMPROPER PADDING PER HOST (FOR TCP PROTOCOL)

| *Date* | *Monday* | *Tuesday* | *Wednesday* | *Thursday* | *Friday* |
|---|---|---|---|---|---|
| No. of frames | 177,653 | 374,285 | 217,099 | 559,866 | 370,579 |

V. POSSIBLE COUNTERMEASURES

Our proposal of the new steganographic system, *PadSteg*, proves that such phenomenon like *interprotocol steganography* is possible and may pose a threat to network security.

In today's LANs, with security measures they provide, *PadSteg* will be hard to detect. The main reason for this is that current IDS/IPS (Intrusion Detection/Prevention System) systems are rarely used to analyze all traffic generated in a LAN as this would be hard to achieve from the performance point of view. Moreover, usually IDSs/IPSs operate on signatures, therefore they require continuous signatures updates of the previously unknown steganographic methods, especially, if the information hiding process is distributed over more than one network protocol (as it is in *PadSteg*).

Thus, the best steps we can take to alleviate *PadSteg* in LANs are to:
- Ensure that there are no NICs with *Etherleak* vulnerability in the LAN.
- Enhance IDS/IPS rules to include *PadSteg* and deploy them in LANs.
- Improve access devices (e.g. switches) by adding active warden functionality [7] i.e. ability to modify (set to zeros) Ethernet frame padding if an improper one is encountered.

Implementation of the specified countermeasures greatly minimizes the risk of successful *PadSteg* utilization.

VI. CONCLUSIONS

In this paper we presented new steganographic system - *PadSteg* – which is the first information hiding solution based on *interprotocol steganography*.

It may be deployed in LANs and it utilizes two protocols to enable secret data exchange: Ethernet and ARP/TCP. A steganogram is inserted into Ethernet frame padding but one must always "look" at the other layer protocol (ARP or TCP) to determine whether it contains secret data or not. Based on the results of conducted experiment the average steganographic bandwidth of *PadSteg* was roughly estimated to be 27 bit/s. It is a quite significant number considering other known steganographic methods.

In order to minimize the potential threat of *interprotocol steganography* to public security identification of such methods is important. Equally crucial is the development of effective countermeasures. This requires an in-depth understanding of the functionality of network protocols and the ways in which they can be used for steganography.

However, considering the complexity of network protocols being currently used, there is not much hope that a universal and effective steganalysis method can be developed. Thus, after each new steganographic method is identified, security systems must be adapted to the new, potential threat.

As a future work larger volumes of traffic from different LANs should be analyzed in order to pinpoint more accurately *PadSteg* feasibility and calculate its steganographic bandwidth.